\documentstyle[12pt]{article}
\begin{document}

\title{
Structural and ultrametric properties of twenty(L-alanine)
}
\author{B. VELIKSON\\
Section de Biophysique des Prot\'eines et des Membranes\\
DBCM, Centre d'Etudes de Saclay,\\
91191 Gif-sur-Yvette Cedex, FRANCE\\
J. BASCLE, T. GAREL and H. ORLAND\\
Service de Physique Th\'eorique, CE-Saclay\\
91191 Gif-sur-Yvette, France\\
}
\maketitle
\begin{abstract}
We study local energy minima of twenty(L-alanine). The minima are
generated using high-temperature Molecular Dynamics and Chain-Growth Monte
Carlo simulations, with subsequent minimization. We find that the
lower-energy configurations are $ \alpha $-helices for a wide range of
dielectric constant values $ (\epsilon  = 1,10,80), $ and that there is no
noticeable
difference between the distribution of energy minima in $ \phi \psi $ space
for
different values of $ \epsilon . $ Ultrametricity tests show that lower-energy
$ (\alpha $
-helical) $ \epsilon =1 $ configurations form a set which is ultrametric to a
certain
degree, providing evidence for the presence of fine structure among those
minima. We put forward a heuristic argument for this fine structure. We also
find evidence for ultrametricity of a different kind among $ \epsilon =10 $
and $ \epsilon =80 $
energy minima. We analyze the distribution of lengths of $ \alpha $-helical
portions
among the minimized configurations and find a persistence phenomenon for the
$ \epsilon =1 $ ones, in qualitative agreement with previous studies of
critical lengths.
\end{abstract}
\newpage
\let\vect=\overrightarrow

\section{INTRODUCTION}

There is a number of reasons for studying the structural properties of
energy minima of poly(L-alanines). We shall cite here just a few:

1. It has been recently confirmed \cite {Head-Gordon}  that
poly(L-alanine) may serve as a good model for many proteins;
namely, that the native backbone geometry for a polypeptide or
protein of M residues very often has  a closely related metastable analog
 in a  poly(L-alanine) of the same number M of
residues. The proteins for which this hypothesis has been successfully checked
include bovine pancreatic trypsin inhibitor (M=58), crambin (M=46),
ribonuclease A (M=124),
and superoxide dismutase (M=158).

2.Our studies of hepta(L-alanine) \cite {us}
showed that all energy minima group themselves into dense
clusters in  configuration space, and the extension of this behavior to
longer molecules is quite plausible and supported, indeed, by the results of
the present paper. On the other hand, there is a general belief that the global
minimum of poly(L-alanine) in the vacuum
is an $\alpha$-helix for sufficiently long molecules. Therefore, we have
reasons to expect that
there are completely or partially $\alpha$-helical clusters in the low-energy
configurations of any
poly(L-alanine), and it would be interesting to find more about the cluster
fine structure, if any.

3. If $\alpha$-helices are in fact the lowest-energy conformations, why is
that so? Is it due mostly to steric or electrostatic effects? What is the
influence of the dielectric constant?

In this paper we study these and related questions, taking as an example
twenty(L-alanine). This molecule is sufficiently long that $\alpha$-helices
become the lowest-energy configurations (contrary to the hepta-alanine case
\cite {us}), but short
enough for an extensive computational study. The energy-minimized
configurations are obtained
by computer simulation and minimized at three different values of dielectric
constant:
$\epsilon=1,10,80$. We find that, indeed, there is  evidence for fine structure
among the $\epsilon=1$ $\alpha$-helical (lowest) minima: the set of such
minima seems {\it ultrametric}. We also find evidence for ultrametricity
(of a different kind)
in the $\epsilon=10,80$ cases. We study
the $\alpha$-helical content of minimized configurations, and find that  for
$\epsilon=1$,
there is  ``persistence'' of an  $\alpha$-helical  structure once it has
reached a
certain length; this is consistent with earlier studies of critical length for
$\alpha$-helix
formation and distribution of $\alpha$-helical portions of longer chains \cite
{Arridge} \cite {PS}
\cite {RipS} \cite {RapS}.  We show that $\alpha$-helices are the
lowest-energy structures independently of $\epsilon$, but that the reason for
this property -- the
relative role of different terms in energy -- does in fact depend on
$\epsilon$. We also note that
the value of the dielectric constant does not much affect the position of
energy minima in the
$\phi\psi$ space: the $\phi\psi$-space clusters of energy minima stay very much
in place.

The paper is organized as follows. In
Section~\ref{modmeth}, we briefly describe the model and  the
sampling methods.
We have generated  Boltzmann-distributed configurations of twenty(L-alanine) at
various temperatures. These configurations were obtained by two
methods: (1) the Langevin Molecular Dynamics simulations run with CHARMM \cite
{Brooks}, \cite
{SmithKarplus} and (2) the
Monte
Carlo Chain-Growth method described in \cite {us}.
Subsequent minimization was done with the Adopted Basis-set
Newton-Raphson  algorithm (ABNR) \cite {Brooks}
under CHARMM.

Section~\ref{res} is dedicated to the
results. In particular, ultrametricity (the evidence for fine
structure among lower-energy minima) is discussed in Section~\ref{ultra}.
The statistics concerning
$\alpha$-helical content among the minima, is presented in
Section~\ref{content}, and the role of various energy terms for
stabilizing the helical conformations is discussed in
Section~\ref{role}.   We end
with a short conclusion.
\vfill
\eject

\section{THE MODEL AND THE METHODS}
\label{modmeth}

In this section we describe the model and
the sampling methodology.

Exactly as we did for hepta(L-alanine) \cite {us},
the 20(L-alanine) molecule is represented as
CO$_2$-CH(CH$_3$)-NH-(CO-CH(CH$_3$)-NH)$_{18}$-CO-CH(CH$_3$)-NH$_3$.
The atom numbering scheme and chemical structure are shown in Fig. 1.

The charges on the carboxyl and ammonium groups are set to zero so as to avoid
dominating charge-charge interactions. In this way the principal
electrostatic interactions will result in hydrogen-bonding between backbone
peptide groups.

\subsection{The energy function}
\label{ener-function}

For both methods mentioned above, we use the same energy
function as described previously \cite {SmithKarplus}, \cite {Reiher}:
$$ E = \sum_{bonds} k_b {(b-b_0)^2}
     + \sum_{angles} k_{\theta}
{(\theta - \theta_o )^2}
     + \sum_{dihedrals} k_\phi (1+cos [n\phi - \delta])
$$
$$      + \sum_{impropers} k_{\nu}(\nu - \nu_o)^2
    + \sum_{i<j} 4 \epsilon_{ij} \Bigl[
 {\left ( {\sigma_{ij} \over {r_{ij}} }\right )} ^{12}
- {\left ({ \sigma_{ij}  \over  r_{ij} }\right )}^6 \Bigr]
     + \sum_{i,j} {332 q_i q_j \over {\epsilon r_{ij}}}\left [ e14(i,j) \right
]
\eqno (1)
$$
where distances are in \AA , angles in radians, and $E$ is in
kcal/mol.

 In Eq. (1) b, $\theta$, $\phi$ and $\nu$ are the
bond lengths, angles, dihedrals and improper torsions
and $b_0$, $\theta _0$, $\nu _0$
are the reference values for these properties.
The force constants associated with these terms are $k_b$,
$k_\theta$, $k_\phi$ and $k_\nu$. For the
intrinsic torsions n is the symmetry number of the rotor
(e.g., 3 for a methyl group) and $\delta$ is the phase angle.

 The nonbonded terms in Eq. (1) are
pairwise additive and consist of a 12-6 Lennard-Jones van der Waals term
and a Coulombic term representing electrostatic interactions between
atomic point partial charges. These nonbonded interactions are included
for 1,4 (vicinal) or higher order atom pairs.
The quantities $r_{ij}$, $\epsilon_{ij}$, $\sigma_{ij}$,
$q_i$ and $q_j$ are the nonbonded distance, the Lennard-Jones well
depth, the Lennard-Jones diameter and the charges for atom pairs
$i$ and $j$.
 The factor $[e14(i,j)]$ in Eq. (1) is 0.5 for (1,4) pairs and 1 for all other
pairs of atoms. A distance-independent relative dielectric constant $\epsilon$
of 1, 10, or 80
is assumed. The parameters used in the energy function can be found, e.g., in
\cite  {us}.

\subsection{Sampling Methods}

We made use of two sampling methods here: (1) traditional Molecular Dynamics,
and (2) Monte Carlo Chain-Growth method \cite {us}. Notice that our aim was
obtaining as many as possible different configurations belonging to
different basins of attraction of the minimizing procedure, rather than
obtaining
good-quality Boltzmann ensembles.  This is why it was
quite appropriate to use finite-temperature configurations produced with
one value of $\epsilon$ as starting points for an energy minimization at
this and {\it other} values of $\epsilon$. Heuristically, this idea was due to
the following consideration: in the approximation in which the molecule's
energy
is dominated by electrostatics, changing $\epsilon$ is equivalent to changing
the temperature.

Changing the value of $\epsilon$ for the purposes of minimization may seem a
rather unusual way of handling a sampling procedure. One may worry that this
could result in an undesired biasing of the ensemble of minima. Such a
consideration would be appropriate if, in doing so, we could {\it lose}
information.  In the case at hands, however, we could certainly only gain in
finding {\it more} minima, wherever they come from. Namely, without using
this procedure, one obtains  the deepest energy
minima at all values of $\epsilon$ in the form of $\alpha$-helices. However,
the statistics is too poor, and one would like to increase it. For this
purpose, we observe that all $(\epsilon=1)$-generated configurations converging
to a
helix upon minimization, belong  to the basins of attraction of
$\epsilon=10$ and $\epsilon=80$ $\alpha$-helical minima as well. In fact,
 this procedure gave us as many $\epsilon=10$ and $\epsilon=80$
$\alpha$-helices as the standard procedure did:
half of those, in our simulation, were
produced by a minimization starting from $\epsilon=1$ strongly-biased
Monte-Carlo generated
configurations.

Thus, in this way
we obtained a better representation of the  minima population for each value of
$\epsilon$ without having to perform more high-cost finite-temperature MC or MD
simulations.

\vskip 1 cm
{\bf(1) MC Chain-Growth Method}

A detailed analysis of the Monte Carlo Chain-growth method can be found in
\cite {us}. In this procedure, an ensemble of chains (molecules) is grown atom
by  atom, so as to generate an ensemble which (in the large-ensemble limit)
obeys the Boltzmann distribution. One advantage of this method is that it does
not assume any particular guess for the initial state but rather builds up the
chain according to Boltzmann weights. Several such Monte Carlo runs at various
temperatures (300K, 1000K,  and 6000K) and values
of $\epsilon$  were performed, giving a population of 86 different
 configurations produced at  $\epsilon=1$, 36 configurations at $\epsilon=10$,
and 16 configurations at $\epsilon=80$.

In one of the $\epsilon=1$ MC simulations, yielding 16 different
configurations,  we used a biased  (guiding field) version of the chain-growth
method \cite {we}. Minimization of these  at all three values of
$\epsilon$ invariably produced completely $\alpha$-helical configurations.

A typical CPU time necessary to produce one configuration was
0.7 s on a Cray 2.

\vskip 1 cm
{\bf(2) Molecular Dynamics}

MD simulations were performed using the program CHARMM \cite {Brooks} with the
Langevin Dynamics algorithm \cite {VanKampen}. In this approach, the equation
of
motion contains a frictional term and a Gaussian external force.
In a one-dimensional representation the equation is as follows,
$$m\ddot x = -U^\prime (x) - \gamma \dot x + f(t),\eqno (12)$$
where the friction coefficient $\gamma$ and the force correlation function
$<f(t)f(0)>=C\delta(t)$ are related by
$C=2\gamma k_B T$. In our simulations, the values of $\gamma/m=40
sec^{-1}$  and $T=1000 K$ were used. We performed two
such simulations, with different starting points ($\alpha$-helix and
$\beta$-sheet), for each of the three values of $\epsilon$. For $\epsilon=1$,
one
simulation of 400 ps and one of 297 ps were  performed. For $\epsilon=10$, the
two runs were 100 ps and 150 ps long. For  $\epsilon=80$, they were 150 ps and
200 ps long. The configurations to be  minimized were taken each 1 ps of the MD
trajectory.

A typical 100-ps MD simulation took the  average of 37 hrs CPU time on an
Alliant VFX-80 machine.

\vskip 1 cm

{\bf(3) Minimization }

As we have already mentioned,
we used configurations produced at each value of $\epsilon$ for minimization
both at that and other $\epsilon$ values.

In this way, we obtained 799 $\epsilon=1$ configurations,
272 $\epsilon=10$ configurations, and 452 $\epsilon=80$ configurations  to be
minimized.
The minimization was done in two stages. To save CPU time, and
because we are more interested in low-lying minima, a preliminary
minimization of 2000
cycles of Adopted Basis-set Newton-Raphson (ABNR) procedure was performed on
all of the above configurations. Then these partly-converged minima were
ordered in energy, and a set of 2-dimensional $\phi\psi$ plots was
obtained.
Then a
number of lower-energy configurations (see Table I) were chosen to be minimized
until the rms energy derivative was $10^{-6}$  kcal/mol $\AA$ or smaller.
(The partly minimized configurations were so close in  configuration
space to actual minima that there was no noticeable difference between the
$\phi\psi$ plots of partially and completely minimized configurations).
This second stage of minimization
took from 3000 to 15000 cycles of the ABNR. Naturally, some configurations
converged to identical minima. The total count of initial and minimized
configurations  is given in Table  I.

Typically, minimization took about 3 hrs per 1000 ABNR steps   CPU time on
an Alliant VFX-80  machine, i.e. 6 hrs for the first, 2000-step stage, with
9 to 45 more hours during the minal minimization stage.

\vfill
\eject

\section{RESULTS}
\label{res}
\subsection{$\phi
\psi$-distribution and classification of minimum-energy
structures}
\label{distr}

The obtained $\phi\psi$-maps are very similar among themselves. There is
not much difference between the maps of different $\phi\psi$ planes, nor
between the maps corresponding to different values of $\epsilon$; we present
two such maps, for $\epsilon=1$ and $\epsilon=80$
(Fig.~2a,~b). All of the maps show clusters of
points located in the same regions as those previously obtained   for
hepta(L-alanine) \cite {us}.
Moreover, we know that their localisation corresponds to
the energy minima of an alanine dipeptide \cite {us}. This means that the
long-distance cross-residue effects do not affect the location of the minima
in the first approximation.

This fact allows us to use the techniques  introduced earlier
in \cite {us}, and characterize a configuration by a set of 18 digits varying
from
1 to 9.
Namely, first we consider the configuration as belonging to the 36-dimensional
space spanned by the 18 inner pairs of $\phi$ and $\psi$ backbone angles.
This makes our study much easier than if we considered the whole (3N-6)=
603-dimensional coordinate space.
Then we divide each $\phi\psi$ plane into rectangles as shown in Fig.~3.  The
division is such that individual clusters of points on the  two-dimensional
maps
(Fig. 2) do not share rectangles. This classification divides the
36-dimensional
space into $9^{18}$ rectangular "boxes", each of  which can be labeled by a
sequence of 18 integers referring to the consecutive $\phi\psi$ planes, e.g. \{
5 6 1 5 7 6 1 1 3 2 9 7 2 4 4 9 6 8 \} etc., so  that we can label all minima
as
belonging to this or that box.  This classification provides an easily
observable characteristic of a  structure, so that, e.g., one can see at a
glance the change that a given configuration undergoes during minimization
(Table II). In our opinion, such a representation is often preferable to
two-dimensional  molecular tracings. While, looking at the latter,  it is quite
difficult to observe that a given structure possesses, say, two
$\alpha$-helical
pieces joined by a non-$\alpha$-helical one, one can immediately see this fact
if
one encounters a structure labeled \{1 2 2 2 2 1 7 2 2 2 2 2 2 2 2 2 2 2\}.
		Moreover, we shall often simplify the notation by writing the latter
structure
as  \{1 2$^{(4)}$ 1 7 2$^{(11)}$\}, like we did in ref. \cite {we}, so that a
complete $\alpha$-helix becomes \{2$^{(18)}$\}.

\subsection{ Characteristics of the low-energy minimized configurations}
\label{char}

For each value of the dielectric constant, we have considered the obtained
energy-minimized configurations.
As we have just said, the $\phi\psi$ distribution of minima is not much
affected by the value of
$\epsilon$. Contrary to the case of hepta(L-alanine),
the lowest-energy configurations always are complete $\alpha$-helices. This is
illustrated in Tables III, IV, V for $\epsilon=1, 10,80$, respectively. We
included in Table III an artificially obtained left-handed
helical structure (3$^{(18)}$). One can see that its energy is much higher (by
about 40 kcal/mol) than those of $\alpha$-helices.

The chain tracing of the lowest-energy $\epsilon=1$ structure  is given in
Fig.~4. The hydrogen bonds are given as broken lines. One should remember that
there is no explicit H-bond term in the CHARMM energy function: H-bonds are
simulated by a combination of the electrostatic and Lennard-Jones terms.

Of the 19  $\epsilon=1$ complete $\alpha$-helices obtained after
minimizing the 799 initial configurations, 5 are the result of
MC simulations, of which 3 were the biased $\epsilon=1$ MC runs, and the
other 2 were nonbiased MC $\epsilon=80$ runs minimized at $\epsilon=1$. The
remaining 14 configurations are the result of $\epsilon=1$
 MD simulations starting from an $\alpha$-helix.

All the 12  $\epsilon=10$ complete $\alpha$-helices obtained after minimizing
the 272 initial configurations are the result
of MC simulations, of which 6 were the biased $\epsilon=1$ MC runs minimized
at $\epsilon=10$, and the other 6 were nonbiased MC $\epsilon=10$ runs.

All the 10  $\epsilon=80$ complete $\alpha$-helices obtained after minimizing
the 452 initial configurations are the result
of MC simulations, of which 5 were the biased $\epsilon=1$ MC runs minimized
at $\epsilon=80$, and the other 5 were nonbiased MC $\epsilon=80$ runs.

The energy differences between the highest and lowest-energy $\alpha$-helices
(2$^{(18)}$ configurations)
are:  11.4 kcal/mol ($\epsilon=1$),
4.1 kcal/mol  ($\epsilon=10$), 2.0 kcal/mol ($\epsilon=80$).

\subsection{Ultrametricity}
\label{ultra}

We find dense clusters of minima in configuration space.
This phenomenon has already been studied for small folded globular proteins
using MD  \cite {ElberKarplus} and MC \cite {NogutiGo} techniques. A thorough
analysis of the clusterization phenomenon was conducted in our previous paper
\cite {us}. It has already been noted more than once (e.g. \cite
{ElberKarplus}) that the
complexity of  the configuration space and the existence of multiple minima
with
small  energy differences bears resemblence to the case of spin glasses. It
comes to mind, therefore, that it would be useful to verify if our distribution
of minimized
conformations shares with spin glasses one of their interesting properties,
i.e. ultrametricity \cite {RTV}.
We remind that a simplistic notion of ultrametricity can be made by imagining
a series of clusters within clusters: they form a tree-like structure, so
that clusters are present
at an arbitrary scale.
A preliminary analysis, for the case of
folded myoglobin, can be found in  \cite {ElberKarplus}, where
no direct evidence for a tree structure
was found.

In our previous paper on hepta(L-alanine) \cite {us},
we looked into the question
whether  the minimized configuration set forms an ultrametric ensemble.
No evidence for such a property was found there.
We are posing the same question for the present case of twenty(L-alanine).
In the $\epsilon=1$ case,
the answer
remains negative, if one considers the
whole set of energy-minimized configurations; and it is positive,
if one restricts the set to
the lower, completely or partially $\alpha$-helical configurations. In the
$\epsilon=10$ and
$\epsilon=80$ cases, the situation is somewhat different.

To test if the 20-alanine distribution is ultrametric, we use the distance
between the minima. There is no unique choice of such a distance. A natural
choice for peptides is the distance in 36-dimensional $\phi \psi$ space,
defined as follows :
$$d=(\sum_1^{18} [(\phi_i-\phi^\prime_i)^2 + (\psi_i-\psi^\prime_i)^2
])^{1/2}$$
corrected to take into account the periodicity of $\phi$ and $\psi$
coordinates:
 if $\vert \phi_i-\phi^\prime_i\vert > 180^\circ$, it is substituted by
$(360^\circ - \vert \phi_i - \phi^\prime_i\vert)$, and the same applies to
$\psi$.

Considering all possible triplets of energy minima, and forming triangles
with vertices at each minimum, one then plots the length of the second longest
side of each triangle against the length of the longest side. In the ideal
ultrametric case, the points will lie on the diagonal $y=x$. If no
ultrametricity
is present, the points will fill the triangle between the diagonal and the line
 $y=x/2$. (The lower bound is a trivial result of the triangle inequality: if
$a,b,c$ are sides of
a triangle, $a<b+c$). If ultrametricity is present to some extent, the density
of points  will
become greater towards the diagonal.

In Figs. 5, 6, and 7 we present such  plots for $\epsilon=1$, $\epsilon=10$,
and $\epsilon=80$ cases. The computations were done for samplings of 133 out
of 799, 69 out of 272, and 61 out of 452 configurations,
respectively (see Table~I).  Namely, in the $\epsilon=1$ and $\epsilon=10$
cases, we took
all fully minimized
configurations for this analysis. In the $\epsilon=80$ case, we felt that 18
fully minimized
configurations would give us too poor a statistics, so that we added 43 upper,
partially minimized
configurations.

In the $\epsilon=1$ case, due to the printer's memory limitations, we could not
present all the $C_{133}^3$=383306 triplets on the graph, so that Fig.~5 shows
a subset of
$C_{67}^3$=47905 triplets generated by taking each second configuration of the
133 configurations
used in computation. The center of inertia of the actual distribution of 383306
points is at
($x=241.0378,\ y= 211.8622$), i.e. on the line $y=0.88x$. The ratio of the
moments of inertia of the
point distribution along the principal axes is 18.0834.

For $\epsilon=10$, the plot consists of
$C_{69}^3$=52394 triplets. Its center of inertia is at ($x=450.6481, \
y=423.6821$), i.e. on
the line $y=0.94x$. The ratio of the moments of inertia of the point
distribution
along the principal axes is 72.107.

For $\epsilon=80$, the plot consists of
$C_{61}^3$=35990 triplets. Its center of inertia is at ($x=599.8134,\ y=
577.8925$), i.e. on
the line $y=0.96x$. The ratio of the moments of inertia of the point
distribution
along the principal axes is 29.005.

These results are to be compared with a similar point distribution generated by
a random
set of points in the 36-dimensional $\phi\psi$ space. In that case, a numerical
analysis
gives the ratio
of the moments of inertia of 4.0067.
This seems to constitute a clear evidence for calling the above distributions
of
lower minima ultrametric to a large degree.

At the same time, we have to stress the difference between the case of
$\epsilon=1$ lower-energy configurations and the two other cases.

As is well known from the theory of spin glasses \cite {RTV}, \cite {Mezard},
one can define roughly two types of ensembles with ultrametric properties:

(i) for the first type (which corresponds to the high-temperature phase in the
spin-glass case), the triangles in the configuration space are mostly
equilateral;

(ii)  the second type (which corresponds to the low-temperature phase in
the  spin-glass case) has, in addition, an isoceles-triangles component.

For $\epsilon = 1$, the 133 lower configurations are mostly $\alpha$-helical
with the third, non-plotted side of the triangle being usually much
smaller than the other two (second type ultrametricity)
. By contrast, for $\epsilon = 10$ and $80$, only
a small fraction of the configurations taken for ultrametricity analysis
are helical, since, as mentioned above, our statistics in these two cases is
rather poor.
In these cases the majority of triangles have three comparable
sides (first type ultrametricity).
To check this point, we considered
separately 495 upper $\epsilon=1$ configurations with no
$\alpha$-helices present at all. In this case, we obtained a plot of
$C_{495}^3$=20092215 triplets of
points, with the center of gravity at ($x=507.1641, \ y=469.5677$) (lying on
the line $y=0.93x$), with
the ratio of the moments of inertia of 21.6; furthermore,
we found that the third non-plotted side of the triangle is comparable
to the first two.
We thus conclude that for all values of
$\epsilon$, there is a first type ultrametricity
among the non-$\alpha$-helical
configurations.

Much more interesting is the ultrametricity among highly-ordered,
$\alpha$-helical configurations
that we see in the  lower-energy $\epsilon=1$ case.  One may try to put forward
a qualitative argument why
ultrametricity could be expected here. Suppose that an independent change in
the $\phi
\psi$ coordinates corresponding, say,  to one turn of the helix can drive the
molecule from one minimum to
another one with a much smaller change in other $\phi\psi$ coordinates.
Consider one particular energy minimum
corresponding to an $\alpha$-helix ( minimum (1) in
Fig.~8). Now consider  another minimum corresponding to a small change in
the position of one turn of
the helix (minimum (2) of Fig.~8).  This minimum  is very close in $\phi\psi$
space  to minimum (1). Still another minimum can be obtained by a small change
in the position of
{\it two} turns of the helix (minimum (3) of Fig.~8). This minimum  is further
away from minimum (1) or (2), but there will be other minima closer to it
obtained from minimum (3) by a small
change in the position of only one turn of the helix (minimum (4) of Fig.~8).
Changing {\it three}
turns of the helix, we obtain a minimum still further away (minimum (5) of
Fig.~8). In this
fashion, we obtain  a hierarchy of clusters (called level-0, level-1, level-2
etc... cluster in
Fig.~8), corresponding to a change inside a given turn of the helix, in one
more turn, in two more
turns etc... Even though this model is necessarily simplistic, and the reality
may be organized in
a somewhat more complicated way, it provides an insight into the reasons for
existence of a fine
structure among this type of configurations.

\vskip 1 cm
\subsection {$\alpha$-helical content: persistence of $\alpha$-helical
structure}
\label{content}

A definition of what is an $\alpha$-helical portion of a secondary structure
differs from one
author to another \cite {ZB}, \cite {LR}, \cite {Daggett}, \cite {RS}.  In our
approach, it is
natural to adopt the definition stemming from our way of labeling
configurations (Section 3.1).
Namely, a configuration is said to have an $\alpha$-helical piece of length $n$
if its label contains
a 2$^{(n)}$ part. We shall also say that such a configuration has an
$\alpha$-helical content of $n$.
The reason for choosing this definition, rather than a definition based on
H-bonds or on explicit
variation of $\phi$ and $\psi$ angles (like in \cite {Daggett}) is that due to
the clustering
phenomenon, all such definitions are equivalent, and ours is much easier to
use.

{\bf (1): $\epsilon=1$ case.}

 We analyzed the relative
distribution of configurations with different $\alpha$-helical content $n$. In
Fig.~9 we show how many $\epsilon=1$
 minimized configurations contain
$\alpha$-helical pieces of given $n$.
(Our convention is such that, e.g.,  a chain labeled  \{2$^{(2)}$ 1 5 7 3 4 8 4
2$^{(3)}$ 3 7 8 2$^{(3)}$\}, which contains one 2-chain of twos and two
3-chains
of twos, will contribute once into the count for 2-chains and once into the
count
for 3-chains. Other conventions could be envisioned.) We see that the
distribution has a deep at $n$=9.

This means that in this  strong electrostatics case  of $\epsilon$=1, we
encounter the phenomenon of persistence of the $\alpha$-helical
structure: once it has reached $n$=10, there is a tendency for the
$\alpha$-helical
chain to become longer, but there is no such tendency for smaller $n$.

These results qualitatively agree with the known existence of a critical
length $l_c$, below
which an $\alpha$-helix is not the lowest-energy state of a polyalanine. Using
a potential different
from ours, an $l_c$ of 16 was found in \cite {Arridge}.

The analysis can be pushed somewhat further.
Dividing all configurations into those which contain $\alpha$-helical pieces
with $n>9$ and those with $n<9$, we plot two energy histograms, marking the
number of configurations of each kind falling in a given energy bin.
These histograms are plotted in
Fig~10. We can see that there
is a
very small overlap of the two histograms: the two kinds of
configurations are quite well separated in energy. Another way to see this
energy separation is to plot the entire energy range spanned by each set of
configurations of a given helical content $n$. This plot is presented in
Fig.~11.

{\bf (2). Other values of $\epsilon$.}

 The helical content behavior differs significantly for other values of
$\epsilon$. The helical length histograms similar to that of Fig.~9 are
given in Figs.~12,13. Since no change in the chain character ever
occurred during the second minimization stage, and in order to increase the
statistics, we took all minimized chains, rather than only the fully minimized
ones, for this analysis. This may have resulted in a multiple count for those
partially minimized chains which after full minimization converged to identical
minima. This does not influence the qualitative character of the histogram.
Here it is especially interesting to look at the helical content distribution
{\it with} and {\it without} taking into account the chains produced by the
biased MC method, which searched specifically for $\alpha$-helices. Unlike in
the $\epsilon=1$ case, where complete $\alpha$-helices ($n$=18) could easily be
found by whichever method, here only very few (6 out of 23) are found by
nonbiased methods for $\epsilon=10$, and 6 out of 24 for $\epsilon=80$.
(Remember that this count
is different from the count of Section~\ref{char}, where all configurations
converging to
identical minima were identified). The helical content distribution falls
rapidly for $\epsilon = 80$,
with some hints of a rise around $n=16$ for $\epsilon=10$.  Nevertheless, the
$n=18$ full
$\alpha$-helices remain the deepest minima in all cases: it remains to find
them. This means that
while the position of the deepest minimum cluster does not change much with
$\epsilon$, its width
falls down rapidly.

Energywise, the deepest minimum found is  -113.285 kcal/mol for $\epsilon=10$,
with reasonably helical configurations running up to -105.599  kcal/mol, and
-68.519
kcal/mol for $\epsilon=80$, with reasonably helical configurations running up
to
-63.525
 kcal/mol.
\vskip 1 cm
\subsection{The role of various energy terms in stabilizing configurations}
\label{role}

The energy components for a sample of 11 lower-energy minimized configurations
are given in Table VI. These configurations are completely or almost
$\alpha$-helical. We can see that the Coulombic interaction is
 about 0.92 of the total energy, and the van der Waals
interaction about 0.09 of the total. The other components can almost be
neglected. Even though the absolute value of energy in CHARMM has no physical
sense and could be
arbitrarily shifted, it is still true that the Coulombic forces give the main
contribution into the
energy differences between configurations (11.74 kcal/mol out of 19,83 kcal/mol
between the lowest
and the highest configurations of Table VI). So in this case, we could think
that the helical shape is
created mainly by electrostatics.

The situation, however, is not that simple: exactly the
same form of the helical lowest minimum is stabilized in case of larger
$\epsilon$ values mostly by
non-electrostatic components.

Looking at Table VII, where we give the same kind of splitting the energy into
components for
$\epsilon=10$ and 80, we can see that in the $\epsilon=10$ case, the energy
difference of
58.03~kcal/mol between the lowest and highest configurations of Table VII, is
due mostly to the
Lennard-Jones forces (45.08 kcal/mol). In the $\epsilon=80$ case, the energy
difference is 8.36
kcal/mol, and consists mostly of Lennard-Jones (2.47 kcal/mol) and dihedrals'
(4.09 kcal/mol)
contributions. The Coulombic contribution is negligeable (.4 kcal/mol). So in
this case, the reason
for obtaining the same shape of the molecule is quite different. In all cases
we can speak about
{\it cooperativity} of various energy components in producing a helical shape.
The Ramachandran plot
for a single residue, with no dipole-dipole interactions, already gives us a
setup for producing a
helix; in the case of strong electrostatics, as the molecule becomes longer,
Coulombic forces
contribute to maintaining an $\alpha$-helical shape.
\vfill
\eject

\section{CONCLUSIONS}

We studied here the energy-minimized structures of twenty(L-alanine).
We confirmed
that the lowest-energy structures are $\alpha$-helices for a wide range of
the dielectric constant. We found, however, that the role of different
energy components in stabilizing these  structures is quite different. At the
same time, the value of $\epsilon$ plays little role in the minima
distribution in the configuration $\phi\psi$ space. Our main results bear upon
the evidence for ultrametric behavior of lowest-energy, $\alpha$-helical
structures and, independently, of higher-energy, disordered structures. We
give a qualitative argument for ultrametricity of lower-energy
configurations.

	We also find, in agreement with previous studies of the critical length for
polyalanine and other polypeptide molecules, that $\alpha$-helical structure
becomes favorable once it has reached a certain length (around ten residues).

\vskip 1cm
We would like to thank J.Smith for useful discussions and suggestions.
\vfill
\eject

\vfill
\eject

%\footnotesize
\begin{center}
\begin{tabular}{|c|c|c|c|c|c|c|c|c|} \hline
{} & \multicolumn{4}{|c|}{partially minimized} &
\multicolumn{2}{|c|}{fully minimized}&taken for & energy range of \\
\cline{2-7}
{} & MD & MC & MC$_\epsilon$ & total & taken for & nonidentical & ultrametr.
&fully minimized \\ [-0.4 mm]
{} & {} & {} & {}            & {}    & minimization & {} & tests      & conf.
(kcal/mol) \\ [-0.4 mm] \hline \hline
$\epsilon=1$ & 697 & 86 & 16 & 799 & 226 & 133 & 133 &-591.202 to -571.371
\\
$\epsilon=10$ & 150 & 36 & 86 & 272 & 84 & 69 & 69 &-116.285 to -53.715 \\
$\epsilon=80$ & 350 & 16 & 86 & 452 & 32 & 18 & 61 & -68.519 to -59.537 \\
\hline
\end{tabular}
\end{center}
\vskip 0.5 cm

\normalsize
Table I. Number  of configurations of 20-alanine obtained during
different simulations. All obtained configurations were at least partially
minimized (2000 cycles of ABNR). Of these, a  lower-energy fraction were fully
minimized.  MC$_\epsilon$ denotes
those MC-generated configurations which were generated at a different value of
$\epsilon$ than the one used during minimization. In the $\epsilon=80$ case,
the 18 fully-minimized nonidentical configurations were complemented by 43
partially-minimized configurations for the ultrametricity test purposes.
\vfill \eject

\begin{tabular}{|l|c|c|} \hline
{} & structure & energy, kcal/mol \\ \hline
initial: &  2 2 2 1 7 6 5 1 1 2 6 6 1 1 5 5 2 6 & -387.2929   \\
minimized: & 2 2 2 1 2 6 5 1 1 1 6 6 1 1 5 1 2 6 & -496.8968  \\ \hline
\end{tabular}
\vskip 0.5 cm
\normalsize
Table II.  An example of a structural change during minimization. The integers
refer to the labeling introduced in Figure 2.

\vfill \eject
\begin{tabular}{|c|c|} \hline
{}& energy \\ [-0.6 cm]
{} & kcal/mol \\ \hline
 2 2 2 2 2 2 2 2 2 2 2 2 2 2 2 2 2 2 & -591.202 \\ [-0.6 cm]
 2 2 2 2 2 2 2 2 2 2 2 2 2 2 2 2 2 2 & -591.099 \\ [-0.6 cm]
 2 2 2 2 2 2 2 2 2 2 2 2 2 2 2 2 2 2 & -590.401 \\ [-0.6 cm]
 2 2 2 2 2 2 2 2 2 2 2 2 2 2 2 2 2 2 & -590.135 \\ [-0.6 cm]
 2 2 2 2 2 2 2 2 2 2 2 2 2 2 2 2 2 2 & -589.997 \\ [-0.6 cm]
 2 2 2 2 2 2 2 2 2 2 2 2 2 2 2 2 2 2 & -589.791 \\ [-0.6 cm]
 2 2 2 2 2 2 2 2 2 2 2 2 2 2 2 2 2 2 & -589.509 \\ [-0.6 cm]
 2 2 2 2 2 2 2 2 2 2 2 2 2 2 2 2 2 2 & -589.437 \\ [-0.6 cm]
 2 2 2 2 2 2 2 2 2 2 2 2 2 2 2 2 2 1 & -589.163 \\ [-0.6 cm]
 2 2 2 2 2 2 2 2 2 2 2 2 2 2 2 2 2 1 & -588.895 \\ [-0.6 cm]
 2 2 2 2 2 2 2 2 2 2 2 2 2 2 2 2 2 2 & -588.827 \\ [-0.6 cm]
 4 2 2 2 2 2 2 2 2 2 2 2 2 2 2 2 2 2 & -588.748 \\ [-0.6 cm]
 4 2 2 2 2 2 2 2 2 2 2 2 2 2 2 2 2 2 & -588.743 \\ [-0.6 cm]
 2 2 2 2 2 2 2 2 2 2 2 2 2 2 2 2 2 2 & -588.690 \\ [-0.6 cm]
...................................&..... \\ [-0.6 cm]
 1 4 2 2 2 2 2 2 2 2 2 2 2 2 2 2 2 1  & -586.820 \\ [-0.6 cm]
...................................&.....  \\ [-0.6 cm]
 1 4 1 7 2 2 2 2 2 2 2 2 2 2 2 2 2 2 & -572.014 \\ [-0.6 cm]
...................................&..... \\ [-0.6 cm]
 3 3 3 3 3 3 3 3 3 3 3 3 3 3 3 3 3 3 & -551.657 \\    [-0.6 cm]
...................................&..... \\ [-0.6 cm]
 1 2 6 2 6 5 1 1 6 6 7 5 7 5 5 1 1 1 & -476.187 \\ \hline
\end{tabular}
\vskip 2 cm
\normalsize
Table III. Minimized configurations, $\epsilon=1$. The left-handed helix
(3$^{(18)}$) has about 40 kcal/mol higher energy than the lowest
$\alpha$-helix.

\vfill
\eject

\begin{tabular}{|c|c|} \hline
{}& energy \\ [-0.6 cm]
{} & kcal/mol \\ \hline
 2 2 2 2 2 2 2 2 2 2 2 2 2 2 2 2 2 2 & -113.285 \\ [-0.6 cm]
 2 2 2 2 2 2 2 2 2 2 2 2 2 2 2 2 2 2  &-112.635  \\ [-0.6 cm]
 2 2 2 2 2 2 2 2 2 2 2 2 2 2 2 2 2 2 & -112.380 \\ [-0.6 cm]
 2 2 2 2 2 2 2 2 2 2 2 2 2 2 2 2 2 2 & -111.798  \\ [-0.6 cm]
 2 2 2 2 2 2 2 2 2 2 2 2 2 2 2 2 2 2 & -111.748 \\ [-0.6 cm]
 2 2 2 2 2 2 2 2 2 2 2 2 2 2 2 2 2 2 & -111.731 \\ [-0.6 cm]
 2 2 2 2 2 2 2 2 2 2 2 2 2 2 2 2 2 2 & -111.555 \\ [-0.6 cm]
 2 2 2 2 2 2 2 2 2 2 2 2 2 2 2 2 2 2 & -111.493 \\ [-0.6 cm]
 2 2 2 2 2 2 2 2 2 2 2 2 2 2 2 2 2 2 & -110.843 \\ [-0.6 cm]
 2 2 2 2 2 2 2 2 2 2 2 2 2 2 2 2 2 2 & -110.135  \\ [-0.6 cm]
 2 2 2 2 2 2 2 2 2 2 2 2 2 2 2 2 2 2 & -109.234 \\ [-0.6 cm]
 2 2 2 2 2 2 2 2 2 2 2 2 2 2 2 2 2 2 & -109.168  \\ [-0.6 cm]
 2 7 2 2 2 2 2 2 2 2 2 2 2 2 2 2 2 2 & -107.925  \\ [-0.6 cm]
 2 7 2 2 2 2 2 2 2 2 2 2 2 2 2 2 2 2 & -106.445 \\ [-0.6 cm]
...................................&..... \\ [-0.6 cm]
 1 7 6 4 2 1 7 2 1 7 2 2 2 2 2 1 7 7 & -100.896 \\ [-0.6 cm]
...................................&..... \\ [-0.6 cm]
 5 2 2 2 2 8 8 7 2 2 2 2 2 2 2 2 3 5 &  -95.879 \\ [-0.6 cm]
...................................&..... \\ [-0.6 cm]
 5 6 1 5 7 6 1 1 3 2 9 7 2 4 4 9 6 8 &  -53.715 \\ [-0.6 cm]
...................................&.....  \\ \hline
\end{tabular}
\vskip 2 cm
\normalsize
Table IV. Minimized configurations, $\epsilon=10$.
\vfill
\eject

\begin{tabular}{|c|c|} \hline
{}& energy \\ [-0.6 cm]
{} & kcal/mol \\ \hline
 2 2 2 2 2 2 2 2 2 2 2 2 2 2 2 2 2 2  &  -68.5190 \\ [-0.6 cm]
 2 2 2 2 2 2 2 2 2 2 2 2 2 2 2 2 2 2  &  -67.8760 \\ [-0.6 cm]
 2 2 2 2 2 2 2 2 2 2 2 2 2 2 2 2 2 2  &  -67.7040  \\ [-0.6 cm]
 2 2 2 2 2 2 2 2 2 2 2 2 2 2 2 2 2 2  &  -67.6130 \\ [-0.6 cm]
 2 2 2 2 2 2 2 2 2 2 2 2 2 2 2 2 2 2  &  -67.5160 \\ [-0.6 cm]
 2 2 2 2 2 2 2 2 2 2 2 2 2 2 2 2 2 2  &  -67.4910 \\ [-0.6 cm]
 2 2 2 2 2 2 2 2 2 2 2 2 2 2 2 2 2 2  &  -67.4890  \\ [-0.6 cm]
 2 2 2 2 2 2 2 2 2 2 2 2 2 2 2 2 2 2  &  -66.8720 \\ [-0.6 cm]
 2 2 2 2 2 2 2 2 2 2 2 2 2 2 2 2 2 2  &  -66.7250  \\ [-0.6 cm]
 2 2 2 2 2 2 2 2 2 2 2 2 2 2 2 2 2 2  &  -66.4860 \\ [-0.6 cm]
 2 2 2 2 2 2 2 2 2 2 2 2 1 2 2 2 2 2  &  -64.8130 \\ [-0.6 cm]
 2 2 2 2 8 7 8 7 2 2 2 2 2 2 2 2 2 7  &  -63.5790  \\ [-0.6 cm]
 2 2 1 7 2 2 2 2 1 7 2 2 2 2 1 8 7 2  &  -61.6840   \\ [-0.6 cm]
 2 2 2 1 7 2 2 2 2 2 2 1 7 2 2 2 1 2  &  -61.5060 \\ [-0.6 cm]
...................................&..... \\ [-0.6 cm]
 1 1 2 6 1 1 1 5 1 6 1 1 1 1 4 2 5 8  &  -52.4166  \\ [-0.6 cm]
 1 2 2 2 2 2 2 2 1 1 5 1 2 1 3 5 5 5  &  -50.0045 \\ [-0.6 cm]
...................................&..... \\ [-0.6 cm]
 2 1 3 2 2 1 2 1 8 7 2 4 9 1 1 1 5 1  &  -35.7261  \\ [-0.6 cm]
...................................&..... \\ [-0.6 cm]
 5 6 1 5 7 6 1 1 3 2 9 7 2 4 4 9 5 8  &  -17.6246 \\ [-0.6 cm]
 5 6 1 5 7 6 1 1 3 2 9 7 2 4 4 9 6 8  &  -17.3373 \\ [-0.6 cm]
...................................&..... \\ \hline
\end{tabular}
\vskip 2 cm
\normalsize
Table V. Minimized configurations, $\epsilon=80$.
\vfill
\eject

\begin{tabular}{|c|c|c|c|c|c|c|}\hline
energy & bonds & angles & dihedrals & impropers & vdW & Coulomb \\ \hline
   -591.202 &    0.277  &   4.070  &   5.567 &    1.160 &  -59.021 & -543.255
\\ [-0.6 cm]
  -591.099  &   0.275   &  3.928   &  5.642  &   1.201  & -58.517 & -543.628\\
[-0.6 cm]
  -590.401  &   0.315   &  4.025   &  5.863  &   1.217  & -58.223 & -543.599\\
[-0.6 cm]
  -590.135  &   0.297   &  3.988   &  5.659  &   1.273  & -57.841 & -543.511\\
[-0.6 cm]
  -589.997  &   0.289   &  3.889   &  5.628  &  1.254  & -58.263 & -542.795\\
[-0.6 cm]
     ... & ... & ... & ... & ... & ... & ... \\
  -572.478  &   0.452  &   4.712  &   6.755   &  1.418 &  -50.313 & -535.501\\
[-0.6 cm]
  -572.430  &   0.445  &   5.197  &   6.878   &  1.634 &  -55.455 & -531.129\\
[-0.6 cm]
  -572.314  &   0.462  &   4.538  &   6.528   &  1.518 &  -55.084 & -530.275\\
[-0.6 cm]
  -572.244  &   0.465  &   4.536  &   6.508   &  1.516 &  -55.013 & -530.256\\
[-0.6 cm]
  -572.014  &   0.446  &   7.293  &   6.650   &  1.662 &  -54.686 & -533.378\\
[-0.6 cm]
  -571.371  &   0.447  &   4.864  &   7.218   &  1.760 &  -54.144 & -531.515\\
\hline \multicolumn{7}{|c|} { values averaged over 133 lower minima:} \\ \hline
  -580.967   &  0.398  &   4.733   &  6.652  &   1.440  & -55.008 & -539.182\\
\hline
\end{tabular}
\vskip 1 cm
Table VI. Energy components for the lower-energy (mostly $\alpha$-helical)
$\epsilon=1$ configurations.

\vfill
\eject

\begin{tabular}{|c|c|c|c|c|c|c|c|}\hline
$\epsilon$ &energy & bonds & angles & dihedrals & impropers & vdW & Coulomb \\
\hline
10 & -111.748  &   0.626  &   5.153 &    4.440  &   0.132 &  -70.719 &  -51.379
\\ [-0.6 cm]
10 & -111.555  &   0.616  &   4.710  &   4.177  &   0.116 &  -70.752 &  -50.421
\\ [-0.6 cm]
10 & -111.493  &   0.560  &   4.739  &   4.278  &   0.122 &  -70.750  & -50.441
\\ [-0.6 cm]
 {} &    ... & ... & ... & ... & ... & ... & ... \\  [-0.6 cm]
10 &  -54.822   &  1.220   &  7.558  &  4.813  &   0.155 & -26.431  & -42.137
\\ [-0.6 cm]
10 &   -54.034  &   1.242  &   7.831 &    5.039 &  0.170 &  -26.148 &  -42.168
\\ [-0.6 cm]
10 & -53.715    & 1.243    & 7.815   &  4.848   &  0.155 &  -25.639 &  -42.137
\\ \hline
80 & -67.491  &   0.635    & 4.988   &  3.791    & 0.086 &  -70.700 &   -6.291
\\ [-0.6 cm]
80  & -68.519   &  0.616   &  5.108  &   4.039   &  0.095  & -72.016  &
-6.361\\  [-0.6 cm]
80 & -67.613   &  0.671   &  5.018    & 4.025  &   0.089  & -71.154  &  -6.261
\\  [-0.6 cm]
 {} &    ... & ... & ... & ... & ... & ... & ... \\  [-0.6 cm]
80 & -60.446    & 0.847  &  5.386    & 5.662    & 0.050    &-66.622   & -5.770
\\ [-0.6 cm]
80 &   -59.537  &   0.997&     6.641 &    8.864 &    0.143 &  -70.464 &
-5.717 \\ [-0.6 cm]
80 &   -59.132  &   0.997&     5.920 &    7.881 &    0.113 &  -68.225 &
-5.817 \\ \hline
\end{tabular}
\vskip 1 cm
Table VII. Energy components for typical lower-energy ($\alpha$-helical)
$\epsilon=10$ and $\epsilon=80$ configurations.

\vfill
\eject

\section*{FIGURE CAPTIONS}

\noindent
Figure 1: Numbering of atoms for 20-alanine.
The period for the intermediate (.....) region is ten atoms
long; one finds, e.g., nitrogen atoms at 33, 43, 53,..,153,...
\vskip 1 cm

\noindent
Figure 2: Energy minima of 20-alanine on a typical
$\phi \psi$ plane: (a) $\epsilon=1 $, (b) $\epsilon=80$
\vskip 1cm

\noindent
Figure 3: $\phi\psi$-plane splitting generating splitting
of the  36-dimensional $\phi\psi$-space into
"boxes".
\vskip 1 cm

\noindent
Figure 4: Chain tracing of the lowest-energy
$\epsilon=1$ configuration. The hydrogen bonds are given as broken
lines.

\vskip 1 cm
\noindent
Figure 5:  Ultrametricity test for 67 out of 133 lower-energy
minima, $\epsilon=1$. The graph consists of 47905 points: for
each triangle in configuration space formed by three energy
minima, the length of the second longest side $d_2$ is plotted
against that of the  longest side $d_1$. In the ideally
ultrametric case, all points would lie on the diagonal $y=x$.
A subset of only 67 minima was chosen for this graph due to
the printer's memory limitations; actual computation was done
for 383306 points generated by 133 minima.
 \vskip 1 cm

\noindent
Figure 6:  Ultrametricity test for 69 lower-energy minima,
$\epsilon=10$. The graph consists of 52394 points: for each
triangle in configuration space formed by three energy minima,
the length of the second longest side $d_2$ is plotted against
that of the  longest side $d_1$. In the ideally ultrametric
case, all points would lie on the diagonal $y=x$.

\vskip 1 cm
\noindent
Figure 7:  Ultrametricity test for 61 lower-energy minima,
$\epsilon=80$. The graph consists of 35990 points: for each
triangle in configuration space formed by three energy minima,
the length of the second longest side $d_2$ is plotted against
that of the  longest side $d_1$. In the ideally ultrametric
case, all points would lie on the diagonal $y=x$.
\vskip 1 cm

\noindent
Figure 8: A sketch of the hierarchy of energy minima in
$\phi\psi$ space  corresponding to a change in a given turn of
the helix, in two turns, etc...
\vskip 1 cm

\noindent
Figure 9: Frequencies of encountering an $\alpha$-helical
portion of length $n$ among the $\epsilon=1$ minimized
configurations. ($n=18$ corresponds to a complete
$\alpha$-helix).

\vskip 1 cm
\noindent
Figure 10: The energy histograms for $\epsilon=1$ minimized
configurations:
the $y$-axis represents the number of
configurations falling into each 3 kcal/mol energy bin.

{\sl dashed line}: configurations with helical content $n<9$.

{\sl solid line}: configurations with helical content $n>9$.

\vskip 1 cm
\noindent
Figure 11: Energies spanned by configurations characterized by a
given helical content $n$.

\vskip 1 cm
\noindent
Figure 12:  Frequencies of encountering an $\alpha$-helical
portion of length $n$ among the $\epsilon=10$ minimized
configurations. ($n=18$ corresponds to a complete
$\alpha$-helix).

\vskip 1 cm
\noindent
Figure 13:  Frequencies of encountering an $\alpha$-helical
portion of length $n$ among the $\epsilon=80$ minimized
configurations. ($n=18$ corresponds to a complete
$\alpha$-helix).
\vfill
\eject

\end{document}